\documentclass[pre,twocolumn,showpacs,preprintnumbers,amsmath,amssymb]{revtex4}

\usepackage{graphicx}
\usepackage{dcolumn}
\usepackage{bm}
\usepackage{psfrag}

\providecommand*{\pop}[2]{\frac{\partial #1}{\partial #2}}
\providecommand*{\dd}{\mathrm{d}}
\providecommand*{\dod}[2]{\frac{\dd #1}{\dd #2}}
\providecommand*{\op}[1]{\hat{#1}}
\providecommand*{\X}{\op\sigma_x}

\providecommand*{\Z}{\op\sigma_z}

\providecommand*{\Hop}{\op H}
\providecommand*{\Tr}{\operatorname{Tr}}
\providecommand*{\Kom}[2]{\left[#1,#2\right]}
\providecommand*{\Ket}[1]{\left|#1\right>}
\providecommand*{\Bra}[1]{\left<#1\right|}

\begin{document}

\preprint{APS/123-QED}

\title{Global and local relaxation of a spin-chain under exact
  Schr\"odinger and master-equation dynamics }

\author{Markus J. Henrich}
\email{henrich@theo1.physik.uni-stuttgart.de}
\author{Mathias Michel}
\author{Michael Hartmann}
\author{G\"unter Mahler}
\affiliation{%
Institute of Theoretical Physics I, University of Stuttgart,
Pfaffenwaldring 57, 70550 Stuttgart, Germany}%
\author{Jochen Gemmer}
\affiliation{
Department of Physics, University of Osnabr\"uck, 49069 Osnabr\"uck, Germany}


\begin{abstract}
We solve the Schr\"odinger equation for an interacting spin-chain
locally coupled to a quantum environment with a specific degeneracy
structure. The reduced dynamics of the whole spin-chain as well as of
single spins is analyzed. We show, that the total spin-chain relaxes
to a thermal equilibrium state independently of the internal
interaction strength. In contrast, the asymptotic states of each
individual spin are thermal for weak but non-thermal for stronger
spin-spin coupling. The transition between both scenarios is found for
couplings of the order of $0.1 \times \Delta E$, with $\Delta E$
denoting the Zeeman-splitting. We compare these results with a master
equation treatment; when time averaged, both approaches lead to the
same asymptotic state and finally with analytical results.
\end{abstract}

\maketitle

\section{\label{sec:level1}Introduction}

Various attempts have been made to account for thermodynamical
behavior of quantum systems
\cite{Neumann1929,Lindblad1983}. Especially the relaxation into an
equilibrium state in a pure quantum world does not seem to be feasible
since Schr\"odinger dynamics is reversible like the classical
Hamiltonian dynamics. By introducing irreversibility into quantum
mechanics one thus has to face all the old difficulties.

One quite successful way to introduce relaxation behavior into quantum
mechanical models is to consider open systems modeled by a quantum
master equation \cite{Breuer,Weiss1999} or the Lindblad formalism
\cite{Kossakowski1972,Lindblad1975,Gorini1976}. In these approaches
the influence of the environmental system enters the
Liouville-von-Neumann equation for the considered system via
incoherent damping terms. To deduce such a closed evolution equation
several approximations are necessary, e.g., the Born-Markov
assumption.

In the context of quantum thermodynamics has recently been
found irreversible behavior in classes of very small
bipartite quantum systems described by a pure Schr\"odinger evolution
only \cite{GemmerOtte2001,GeMiMa2004}. This approach does not need
those specific assumptions about the environment. Instead, system and
environment are treated as a whole. Even in a small bipartite
systems, consisting of a two level system (``gas system'') coupled
to an environment (``container'') of no more than some hundred levels
thermodynamical behavior is generic -- relaxation occurs to a
theoretically predicted equilibrium state (see \cite{BorowskiGemmer2003}).

In the above mentioned scenarios the gas system has been very small
(from two to five levels) and coupled to an environment without any
structure or selectivity. In a more complex situation the system under
consideration (gas system) could be constructed from several identical
subsystems, e.g.\ the system could be a spin chain. For
such bipartite systems with increased internal complexity the effect of the
coupling topology is not yet completely understood: If we couple a
chain of identical subsystems at one edge only to a quantum
environment as before, the question arises whether the whole system
will still relax into a thermal equilibrium state. Furthermore, we are
interested in whether and when the individual subsystems are also in a
thermal equilibrium state.

According to recent investigations on local temperature of modular
systems \cite{Hartmann2004III,Hartmann2004IV}, we expect the same
global as well as local temperature in the system for weakly coupled
chains. In cases of a stronger coupling this may no longer be the
case (as shown by an open system approach \cite{Michel2003}).

Those chain systems coupled to a quantum environment will be
treated as a closed system subject to Schr\"odinger dynamics and will
be compared to the modeling via a quantum master equation. In
particular, we will analyze chains with different coupling types and
strengths with respect to their local as well as global thermal or
non-thermal properties.

\section{Theoretical Background}
\subsection{\label{sec:level2}The Considered System}

We consider chains of three identical quantum
systems described by the Hamiltonian
\begin{equation}
  \label{eq:m1}
  \Hop_{\text{s}}=\sum_{\mu=1}^3 \Hop_{\text{loc}}(\mu)
  +\frac{\lambda}{I}\sum_{\mu=1}^2 \op I_{\text{s}} (\mu,\mu+1)\;.
\end{equation}
Here the first sum contains the local Hamilton operators
$\Hop_{\text{loc}}(\mu)$ of site. In our case the local Hamiltonian of
site $\mu$ reads
\begin{equation}
\Hop_{\text{loc}} (\mu)=\op 1+\frac{1}{2}\Z(\mu), \qquad \mbox{with}
\quad \mu=1,2,3 
\label{eq:2}
\end{equation}
where $\Z$ denotes the Pauli spin operator. Here and in the following
all energies are taken in units of the Zeeman splitting. The second
sum refers to the next neighbor couplings between the subsystems
normalized by
\begin{equation}
I=\frac{1}{n}\sqrt{\Tr\{ \op I_{\text {s}}^2\}} \;,
\label{eq:5}
\end{equation}
where $n$ is the Hilbert space dimension of the chain. Thus, it is
possible to control the internal coupling strength by the single parameter
$\lambda$ (see (\ref{eq:m1})) only, irrespective of the type of
coupling. In the following we compare two different
interactions -- a random coupling
\begin{equation}
  \op I^{\text {r}}_{\text{s}} (\mu,\mu+1)=\sum_{i=1}^3 \sum_{j=1}^3
  p_{ij} \op \sigma_i(\mu) \otimes \op \sigma_j(\mu+1), 
\label{eq:3}
\end{equation}
where the $p_{ij}$ are normally distributed random numbers in the
interval $[-1,1]$, and a Heisenberg interaction,
\begin{equation}
  \op I^{\text {H}}_{\text {s}} (\mu,\mu+1)=\sum_{i=1}^3 \op
  \sigma_i(\mu) \otimes \op \sigma_i(\mu+1)\;.
\label{eq:4}
\end{equation}

\subsection{\label{sec:level22}The Environment}

\begin{figure}[htbp]
\psfrag{S1}{\tiny spin 1}
\psfrag{S2}{\tiny spin 2}
\psfrag{S3}{\tiny spin 3}
\psfrag{U}[][]{\tiny environment}
\psfrag{s10}{\tiny $E_1^{\text{s}_3}$}
\psfrag{s11}{\tiny $E_0^{\text {s}_3}$}
\psfrag{s20}{\tiny $E_1^{\text {s}_2}$}
\psfrag{s21}{\tiny $E_0^{\text {s}_2}$}
\psfrag{s30}{\tiny $E_1^{\text {s}_1}$}
\psfrag{s31}{\tiny $E_0^{\text {s}_1}$}
\psfrag{n1}{\tiny $N_0(E_0^{\text {e}})$}
\psfrag{n2}{\tiny $N_1(E_1^{\text {e}})$}
\psfrag{n3}{\tiny $N_2(E_2^{\text {e}})$}
\psfrag{n4}{\tiny $N_3(E_3^{\text {e}})$}
\psfrag{n5}{\tiny $N_4(E_4^{\text {e}})$}
\psfrag{n6}{\tiny $N_5(E_5^{\text {e}})$}
\psfrag{n7}{\tiny $N_6(E_6^{\text {e}})$}
\psfrag{n8}{\tiny $N_7(E_7^{\text {e}})$}
\psfrag{t}{\tiny $t[\frac{\hbar}{\Delta E^{\text {s}}}]$}
\includegraphics[width=.4\textwidth]{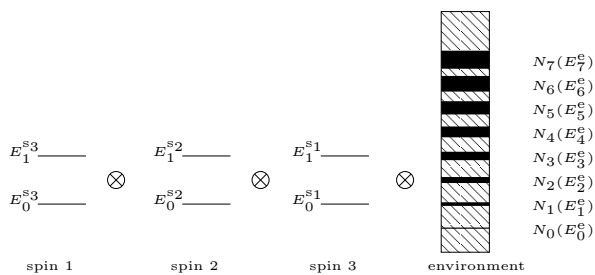}
\caption{\label{fig:System-Bad}Coupling topology of the spin chain to
  the quantum environment.}
\end{figure}
The model system described in the last section, is now coupled to an
adequate environment:
\begin{equation}
  \op H_{\text{tot}} = \op H_{\text{s}} + \op H_{\text{e}}
                     + \frac{\kappa}{I_{\text{se}}} \op I_{\text{se}}\;.
\label{eq:9}
\end{equation}
The first term is the Hamiltonian of the chain, $\op H_{\text{e}}$
refers to the Hamiltonian of the environment and $\op I_{\text se}$ is
the interaction between system and environment.
After normalizing $\op I_{\text{se}}$ via
$I_{\text{se}}=\frac{1}{n}\sqrt{\Tr\{ \op I_{\text {se}}^2\}}$
(compare (\ref{eq:5})) the external coupling strength is controlled by
the parameter $\kappa$ only. This system environment coupling is taken
to be small in all cases as a precondition to allow for thermodynamic
behavior. The interaction will be chosen in such a way that it couples
only to one boundary of the system (see Fig. \ref{fig:System-Bad}),
say to the third spin of the chain. Furthermore, the interaction
should allow for energy exchange between system and environment.

In the following we will consider both, a complete solution of the
Schr\"odinger equation for system and environment according to $\op
H_{\text{tot}}$ as well as a quantum master equation, simulating a
corresponding bath coupling.

\subsection{\label{sec:level23}Hilbert space average approach}
Let us start with the complete model under Schr\"odinger dynamics.
At first sight it may not seem clear how a totally time reversible
equation like the Schr\"odinger equation could produce something like
irreversible behavior. To clarify this point we introduce some aspects
of quantum thermodynamics. A complete derivation of all aspects of
this theory is beyond the scope of this article. Therefore we only sum
up some central aspects of this approach and refer the interested
reader to \cite{Gemmer2003,GeMiMa2004}.

According to quantum thermodynamics (see \cite{GeMiMa2004}), the
system proper relaxes to a Gibbsian state whenever the density of
states of the environment is an exponential function of energy; the
latter is typical for many body systems. We model such a quantum
environment here by a system (Fig. \ref{fig:System-Bad}, the width
of the energy levels should indicate the degeneracy) of the eight
energy levels $E_i^{\text {e}}$ with degeneracies given by
\begin{equation}
N^{\text {e}}(E_i^{\text {e}})=N_0^{\text {e}} 2^{E_i^{\text {e}}}\;.
\label{eq:7}
\end{equation}
This may seem rather artificial, but the idea is as follows: the
environment may possess a continuous spectrum, but due to the weak
coupling between s and e the system only couples to the resonant
levels. Thus we can neglect all the other levels. This environment is
then expected to induce on the spin-chain a canonical state with the
reciprocal temperature $\beta$,
\begin{equation}
\frac{1}{k_{\text{B}} T}=\beta=\pop{}{E} \ln N^{\text {e}}(E)
\label{eq:8}
\end{equation}
where $k_{\text{B}}$ is the Boltzmann-constant.
With the special degeneracy structure (\ref{eq:7}) this leads to $\beta=\ln 2$.

For a numerical test and to avoid any bias we pick for $\op
I_{\text{se}}$ a hermitian random matrix from an ensemble with the
distribution \cite{Mehta1991} 
\begin{equation}
P(\op I_{\text{se}})=\frac{1}{\sqrt{4 \pi}} \text{e}^{-\frac{\Tr
    \left\lbrace \op I_{\text{se}}^2\right\rbrace }{4}}.
\label{eq:6}
\end{equation}

%
%
\emph{Theoretical Predictions:}
In contrast to the master equation approach, the dynamics of the
closed system (spin chain and environment) is purely Schr\"odinger-type.
The von Neumann entropy of the completely closed system thus remains constant.
But by splitting up the whole system into two parts, a small system
taken to be the spin-chain and a large one, the environment, the von
Neumann entropy of the parts can change in time. As a matter of fact,
the small system shows a thermodynamical behavior if two restrictions
are met:
\begin{itemize}
\item[a)] The coupling between the small system and the environment
  should be small, i.e.\ the energy contained in the interaction has
  to be much smaller than the local energies,
\begin{equation}
\left\langle \op I_{\text {se}} \right\rangle \ll \left\langle
  \Hop_{\text {s}} \right\rangle ,  \left\langle \Hop_{\text {e}} \right\rangle
\label{eq:17}
\end{equation}
which guarantees that the spectrum of the environment is not disturbed
too much.
\item[b)] The Hilbert space of the environment should be very large
  compared to the Hilbert space of the small system.
\end{itemize}
Conversely, all systems meeting these conditions may be called
thermodynamical.

The concrete form of the interaction defines an accessible region
within the whole Hilbert space, selected by the initial state. Using a
model with full energy exchange between system and environment
(canonical situation) only one supplementary condition remains active,
the overall energy conservation defining the accessible region. By a
topological investigation of the Hilbert space (see
\cite{GeMiMa2004}), the details of which are beyond the scope of this
article, it is possible to show that the state of the complete system
will enter a very large region (the dominant region) within the
accessible region, for which the system under consideration is in a
state with approximately maximum von~Neumann entropy. The respective
energy distribution of the small system in this dominant region, i.e.\ the
probability to find the small system in an energy eigenstate
$E^{\text{s}}_i$ (no degeneracy) for an initial state with a sharp
energy is then given by
\begin{equation}
  W^{\text{d}}(E^{\text{s}}_i)
  = \frac{N_i(E^{\text{e}}_i)}{N_{\text{tot}}}\;,
\label{eq:18}
\end{equation}
(for a complete derivation of (\ref{eq:18}) see \cite{Gemmer2003, GeMiMa2004}).
$E^{\text{s}}_i$ is the i-th energy level of the spin system,
$E^{\text{e}}_i$ the corresponding energy level in the environment
with the degeneracy $N_i(E^{\text{e}}_i)$ and
$N_{\text{tot}}=\sum_{i=1}^8 N_i(E^{\text{e}}_i)$ is the total number
of levels in the environment.

%
%
\subsection{\label{sec:level24}Master equation approach}

Master equation approaches describe the dynamics of open systems. They
have been widely applied to describe system bath models, in particular
in quantum optics \cite{Breuer}. Their derivation is standard and can
be found in several textbooks. However, since our system, the spin
chain, has an internal structure and since only one of the boundary
spins directly couples to the bath, special care must be taken:

The entire dynamics of the system coupled to the bath is given by the
Liouville-von-Neumann equation, from which the Nakajima-Zwanzig
equation can be derived \cite{Breuer}. Assuming that the bath is in a
thermal state,
\begin{equation}
\rho_{\text {e}} = \frac{\text{e}^{- \beta H_{\text {e}}}}{Z_{\text {e}}}\,
\label{eq:42}
\end{equation}
an expansion of the latter up to second order in the
system bath coupling reads:
\begin{equation}\label{eq:23}
\dod{\rho_\text s}{t}=-i\Kom{\op{H}_\text s}{\rho_\text s}
- \int_0^t dt' \, \text{Tr}_{\text {e}} \Kom{
  \hat{I}_{\text{se}}}{\Kom{ \hat{I}_\text{se} (t'-t)} 
{\rho_{\text s} \otimes \rho_{\text e}}}\, ,
\end{equation}
where $\rho_\text s$ the reduced density matrix of the spin-system and
$\hat{I}_\text{se} (t)$, the system bath interaction in the
interaction picture reads,
\begin{equation}
\hat{I}_\text{se} (t) = \text{e}^{i (\hat{H}_\text s + \hat{H}_{\text
    {e}}) t} \, \hat{I}_\text{se} 
\text{e}^{- i (\hat{H}_\text s + \hat{H}_{\text {e}}) t} \, ,
\label{eq:41}
\end{equation}
the time dependence of which can be computed in the eigenbasis of
$\hat{H}_\text s$ and $\hat{H}_{\text {e}}$. Applying, as usual, the
Markov approximation, the integral of (\ref{eq:23}) can be computed
and the following form of the damping rates is found:

\begin{equation}
\dod{\rho_\text s}{t}=-i\Kom{\Hop_\text s}{\rho_\text s}+
(\op A \op \Gamma^a \rho_\text s)-(\op \Gamma^a \rho_\text s \op
A)+(\rho_\text s \op \Gamma^b \op A)-(\op A \rho_\text s \op
\Gamma^b)\, ,
\label{eq:11}
\end{equation}
where
\begin{equation}
\op A=\op 1(1) \otimes \op 1(2) \otimes \X (3).
\label{eq:16}
\end{equation}
As can be seen the environment has been locally coupled to spin~3 only.

Denoting the energy eigenvalues and eigenvectors of the spin-system by
$E_{\text{s}}^i$ and $\Ket i$, respectively 
\begin{equation}
\Hop_{\text{s}} \Ket i= E_i^\text s \Ket i
\label{eq:40}
\end{equation}
and defining $\omega_{ij}=\hbar (E^{\text{s}}_i - E^{\text{s}}_j)$,
the transition matrices $\Gamma^{a/b}$ have the following matrix
representation in the eigenbasis of $H_{\text{s}}$: 
\begin{equation}
\Bra i \Gamma^{a} \Ket j=\kappa
\frac{1}{\text{e}^{\omega_{ij}\beta_{\text {e}}-1}}A_{ij}, \qquad
\mbox{for} \quad \omega_{ij} > 0 
\label{eq:12}
\end{equation}
\begin{equation}
\Bra i \Gamma^{a} \Ket j=\kappa
\frac{\text{e}^{\omega_{ji}\beta_{\text
      {e}}}}{\text{e}^{\omega_{ji}\beta_{\text {e}}-1}}A_{ij}, \qquad
\mbox{for} \quad \omega_{ij} < 0 
\label{eq:13}
\end{equation}
\begin{equation}
\Bra i \Gamma^{b} \Ket j=\kappa
\frac{\text{e}^{\omega_{ij}\beta_{\text
      {e}}}}{\text{e}^{\omega_{ij}\beta_{\text {e}}-1}}A_{ij}, \qquad
\mbox{for} \quad \omega_{ij}>0 
\label{eq:14}
\end{equation}
\begin{equation}
\Bra i \Gamma^{b} \Ket j=\kappa
\frac{1}{\text{e}^{\omega_{ji}\beta_{\text {e}}-1}}A_{ij}, \qquad
\mbox{for} \quad \omega_{ij}<0 
\label{eq:15}
\end{equation}
Note that by virtue of (\ref{eq:41}) the real damping rates of
(\ref{eq:12}) - (\ref{eq:15}) only appear if the transition matrices
$\Gamma^{a/b}$ are represented in the eigenbasis of $\Hop_\text
s$. Therefore, simply writing down Lindblad damping terms
\cite{Lindblad1976} for spin~3, the one which directly couples to the
bath, would in general, lead to wrong results
\cite{Saito2000}. However, if the coupling between the spins in the
chain is small enough, additional approximations can be made and one
will obtain the Lindblad type damping rates for spin~3 \cite{Capek2000}.

It should also be mentioned, that the thermal state
\begin{equation}
\rho_\text s = \frac{\text{e}^{- \beta \Hop_\text s}}{Z_\text s}
\end{equation}
is the stationary and therefore asymptotic solution of (\ref{eq:11}).

\section{\label{sec:level4}Numerical Results}
\subsection{\label{sec:level4a} Spectral-Temperature}
We will use as a measure to characterize the asymptotic state of the
spin system and the single spins a "spectral-temperature" defined as
\begin{equation}
\begin{split}
\frac{1}{k_{\text {B}} T}:=-\left (1-\frac{W_0+W_M}{2} \right)^{-1} \hspace{2cm}\\
\sum_{n=1}^M\left( \frac{W_n+W_{n-1}}{2}\right)\frac{\ln W_n-\ln
  W_{n-1}}{E_n-E_{n-1}}\, ,
\end{split}
\label{eq:21}
\end{equation}
where $W_n$ is the probability to find the system at the $n$-th energy
level $E_n$, $M$ is the highest and $0$ the lowest energy level. The
main idea here is to assign a Boltzmann factor to each pair of
neighboring energy levels. The spectral-temperature then simply is the
average over all these factors weighed by the corresponding occupation
probabilities. This spectral-temperature 
also exists in non-equilibrium situations but coincides with the
standard definition of temperature (e.\ g.\ \cite{LL5}) only for a
canonical state.

For the calculation of the spectral-temperatures according to
(\ref{eq:21}) we insert for $W_n$ the long time average $\left\langle
  \rho^{nn}_\text s (t) \right\rangle _t$ of the respective occupation
probability.

\subsection{\label{sec:level41}Schr\"odinger dynamics}
To demonstrate the relaxation behavior of the spin-chain under
Schr\"odinger dynamics we have diagonalized the total Hamiltonian
$\Hop_{\text {tot}}$ (see (\ref{eq:6})) and solved the exact
Schr\"odinger equation 
\begin{equation} 
\dod{}{t}\Ket{\psi_{\text {tot}}(t)}=- \frac{i}{\hbar}\Hop_{\text
  {tot}}\Ket{\psi_{\text {tot}}(t_0=0)}. 
\label{eq:20}
\end{equation}

\begin{figure}[htbp]
\psfrag{W}{\tiny $\rho^{nn}_{\text {s}}(t)$}
\psfrag{t}{\tiny $t[\frac{\hbar}{\Delta E^{\text {s}}}]$}
\includegraphics[width=.3\textwidth,angle=270]{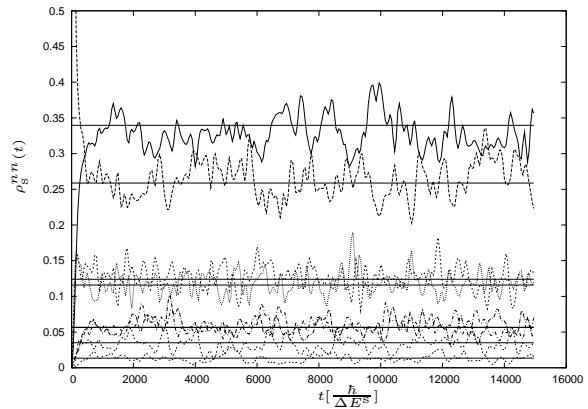}
\caption{\label{fig:System-Sdyn-Z} Relaxation of the spin chain s into
  equilibrium under Schr\"odinger dynamics for $\lambda=0.4$ and
  $\kappa=0.001$ with a random interaction. The horizontal lines are
  the expected probabilities from (\ref{eq:18}) to find the system at
  the corresponding energies. These are in good accordance with the
  time-averaged values of $\rho_{\text {s}}(t)$. $\lambda$ and
  $\kappa$ are energies taken in units of the Zeeman splitting. (Note
  that $W^d(E^{\text s}_4)=W^d(E^{\text s}_5)$ (\ref{eq:18}).)}
\end{figure}

\begin{figure}[htbp]
\psfrag{W}{\tiny $\rho^{nn}_{\text {s}}(t)$}
\psfrag{t}{\tiny $t[\frac{\hbar}{\Delta E^{\text {s}}}]$}
\includegraphics[width=.3\textwidth,angle=270]{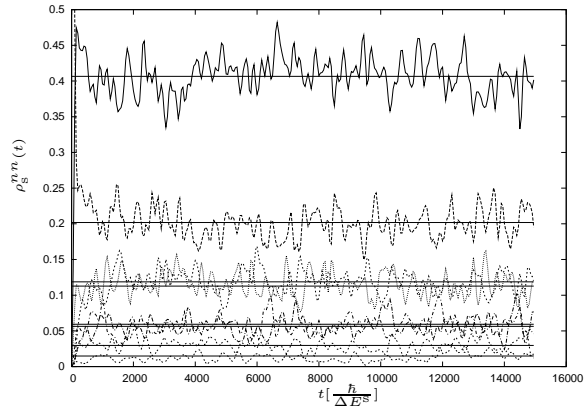}
\caption{\label{fig:System-Sdyn-H}Relaxation into equilibrium under
  Schr\"odinger dynamics as of Fig.~\ref{fig:System-Sdyn-Z} but with a
  antiferromagnetic Heisenberg interaction.}
\end{figure}

The initial state $\Ket{\psi_{\text {tot}}(t_0)}$ has been taken as a
product state of the spin chain and the environment.
From the density matrix $\rho_{\text
  {tot}}(t) = \Ket{\psi(t)} \Bra{\psi(t)}$ we trace out the
environment and transform the resulting reduced density matrix into
the eigenbasis of the spin-system. The time evolution of this reduced
density matrix $\rho_\text s(t)$ is plotted in
Fig.~\ref{fig:System-Sdyn-Z} for an internal random coupling and in
Fig.~\ref{fig:System-Sdyn-H} for a Heisenberg coupling. (We restrict
our analysis to $\lambda > 0$, the antiferromagnetic
case. Corresponding results can be obtained for $\lambda < 0$.)

Both evolutions have been obtained with the coupling parameters
$\kappa=0.001$ and $\lambda=0.4$. The horizontal lines are the
probabilities as expected from (\ref{eq:18}). As can be seen, for both
internal couplings, the spin-system relaxes into a state which is in
accordance with (\ref{eq:18}). The probability fluctuations can be
interpreted as a finite size effect of the environment.

\begin{figure}[htbp]
\psfrag{T}{\tiny $k_{\text {B}} T[\Delta E]$}
\psfrag{lambda}{\tiny $\lambda[\Delta E^{\text {s}}]$}
\psfrag{exp}{\tiny $T^{\text {e}}$}
\psfrag{1}{\tiny $T^{\text {loc}}_1$}
\psfrag{2}{\tiny $T^{\text {loc}}_2$}
\psfrag{3}{\tiny $T^{\text {loc}}_3$}
\psfrag{spec}{\tiny $T^{\text {s}}$}
\includegraphics[width=.3\textwidth,angle=270]{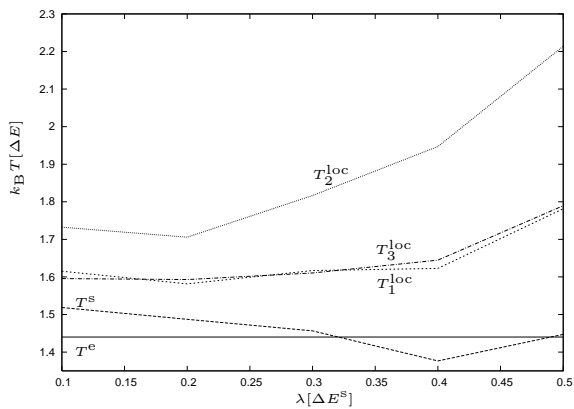}
\caption{\label{fig:Temp-Schr-Z} Global and local temperatures as a
  function of $\lambda$ for a random interaction.}
\end{figure}

\begin{figure}[htbp]
\psfrag{T}{\tiny $k_{\text {B}} T[\Delta E]$}
\psfrag{lambda}{\tiny $\lambda[\Delta E^{\text {s}}]$}
\psfrag{s}{\tiny $T^{\text {s}}$}
\psfrag{1}{}
\psfrag{2}{\tiny $T^{\text {loc}}_2$}
\psfrag{3}{\tiny $T^{\text {loc}}_1= T^{\text {loc}}_3$}
\psfrag{exp}{\tiny $T^{\text {e}}$}
\includegraphics[width=.3\textwidth,angle=270]{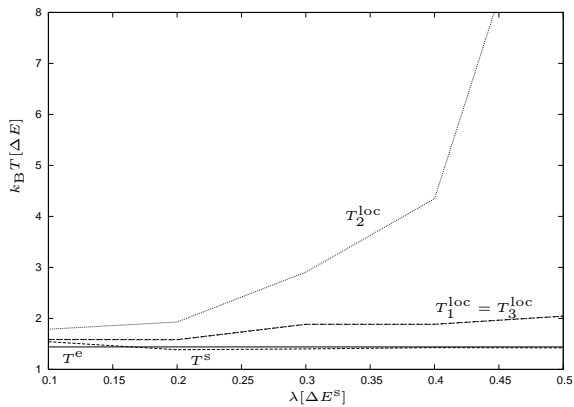}
\caption{\label{fig:Temp-Schr-H} Global and local temperatures as a
  function of $\lambda$ for a antiferromagnetic Heisenberg
  interaction.}
\end{figure}

\emph{Global and local Temperatures:}
Now that we have demonstrated the relaxation of the spin-chain into a
thermal equilibrium state, we want to analyze the relaxation behavior
of the spin-chain depending on the internal interaction strength
$\lambda$. The approach of \ref{sec:level23} is independent of the
internal structure of the small system. Thus the relaxation behavior
should be independent of $\lambda$ and the temperature of the
spin-chain should be identical to that induced by the environment
$T^{\text {e}}=\frac{1}{\ln 2}$ (see (\ref{eq:7})). To specify the
temperature of the spin-chain $T^{\text {s}}$ we use the
"spectral-temperature" defined in (\ref{eq:21}). We compare the
spectral-temperature of the total spin-chain to those of each
individual spin. For the local spectral-temperature $T^{\text
  {loc}}_\mu$ of a single spin, (\ref{eq:21}) reduces to
\begin{equation}
T^{\text {loc}}_\mu=-\frac{E_1^\mu-E_0^\mu}{\text{ln}(W_1^\mu)-\text{ln}(W_0^\mu)},
\label{eq:24}
\end{equation}
where $W^\mu_i$ is the long time average of the probability to find
the $\mu$-th spin at the energy $E^\mu_i$. These quantities are
plotted as a function of $\lambda$ for a random interaction in
Fig.~\ref{fig:Temp-Schr-Z} and for an antiferromagnetic Heisenberg
interaction in Fig.~\ref{fig:Temp-Schr-H}.

The spectral-temperature of the spin chain $T^{\text {s}}$ approaches
the temperature $T^{\text {e}}$ imposed by the environment,
irrespective of the internal coupling strength $\lambda$. However, the
spectral-temperatures of each single spin is found to increase with
increasing $\lambda$. The reason for this behavior is that the local
spin-system is disturbed more and more with increasing $\lambda$ although the
whole spin-system continues to reach a canonical state \cite{Hartmann2004IV}.

Especially for a Heisenberg coupled spin chain one can verify
analytically that for a canonical state with some respective
temperature, the local temperatures deviate more and more with
increasing $\lambda$ (see Sec. \ref{sec:analytic}). Note, that for a
ferromagnetic Heisenberg coupling the spectral-temperatures of each
spin show a different behavior: they decrease with increasing
$\lambda$. In any case, they deviate from $T^\text e$.

To verify that the state of the total spin-chain $\rho_\text s$ is
indeed canonical we test the off-diagonal elements. All absolute
values are smaller than $10^{-4}$. Therefore we argue that the state
of the total spin-chain is indeed a canonical one for all practical
purposes and its spectral-temperature can be identified as the
thermodynamic temperature.

\begin{figure}[htbp]
\psfrag{Di}{\tiny $C$}
\psfrag{lambda}{\tiny $\lambda[\Delta E^{\text {s}}]$}
\includegraphics[width=.3\textwidth,angle=270]{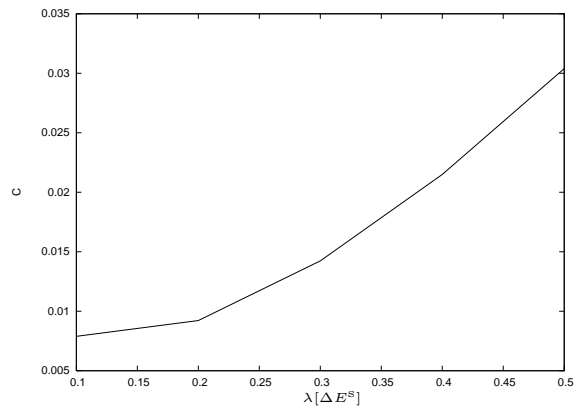}
\caption{\label{fig:Corr-gl-Z} Correlation $C$ (\ref{eq:22}) as a
  function of $\lambda$ for a random interaction.} 
\end{figure}

\begin{figure}[htbp]
\psfrag{Ci}{\tiny $C$}
\psfrag{lambda}{\tiny $\lambda[\Delta E^{\text {s}}]$}
\includegraphics[width=.3\textwidth,angle=270]{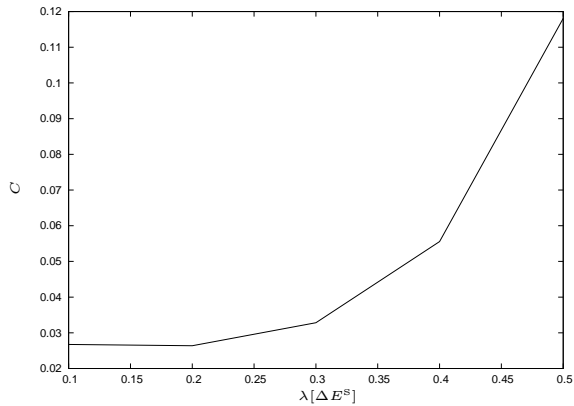}
\caption{\label{fig:Corr-gl-H} Correlation $C$ (\ref{eq:22}) as a
  function of $\lambda$ for a Heisenberg interaction.} 
\end{figure}

The deviation of the local spectral-temperatures from the temperature
of the whole spin-chain can also be understood by analyzing the
correlation $C$ between the spins
\begin{equation}
C=\Tr{\left\lbrace \left[ \left\langle \rho_1 (t) \right\rangle_t
      \otimes \left\langle\rho_2  (t)\right\rangle_t \otimes
      \left\langle \rho_3  (t)\right\rangle_t- \left\langle \rho_\text
        s (t)\right\rangle _t \right] ^2 \right\rbrace}.
\label{eq:22}
\end{equation}
The brackets $\left\langle \right\rangle _t$ denote the time average
of the density matrices of the single spins ($\rho_\mu$) and the
spin system ($\rho_\text s$). Due to increasing $\lambda$ the spins
are more and more correlated. This correlation causes an increase of
the local entropy which leads to an increased spectral-temperature of
each individual spin. Fig. \ref{fig:Corr-gl-Z} shows $C$ as a function
of $\lambda$ for a random interaction and Fig. \ref{fig:Corr-gl-H} for
a antiferromagnetic Heisenberg interaction.


As can be seen from both figures, the correlations $C$ increase with
increasing $\lambda$. The ferromagnetic Heisenberg coupling also show
increasing correlations in the considered ranges of $\lambda$ and
$T^{\text{e}}$.

\subsection{\label{sec:level42}Master equation}
\begin{figure}[htbp]
\psfrag{W}{\tiny $\rho_{\text {s}}^{nn}(t)$}
\psfrag{t}{\tiny $t[\frac{\hbar}{\Delta E^{\text {s}}}]$}
\includegraphics[width=.3\textwidth,angle=270]{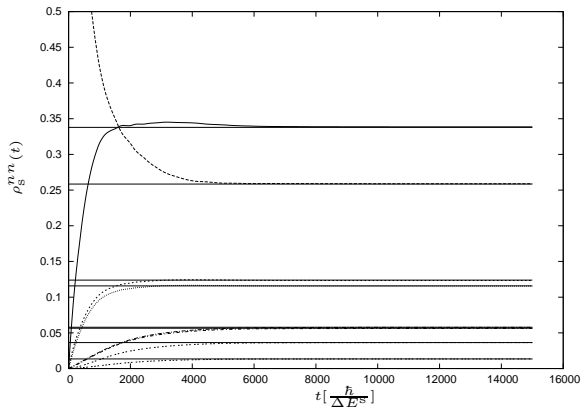}
\caption{\label{fig:System-Master-Z} Time evolution of $\rho^{\text
    {s}}$ under a master equation with random interaction. The
  equilibrium reached is the same as that under Schr\"odinger dynamics
  (Fig. \ref{fig:System-Sdyn-Z}) and the one predicted by
  (\ref{eq:18}) (horizontal Lines).}
\end{figure}

\begin{figure}[htbp]
\psfrag{W}{\tiny $\rho_{\text {s}}^{nn}(t)$}
\psfrag{t}{\tiny $t[\frac{\hbar}{\Delta E^{\text {s}}}]$}
\includegraphics[width=.3\textwidth,angle=270]{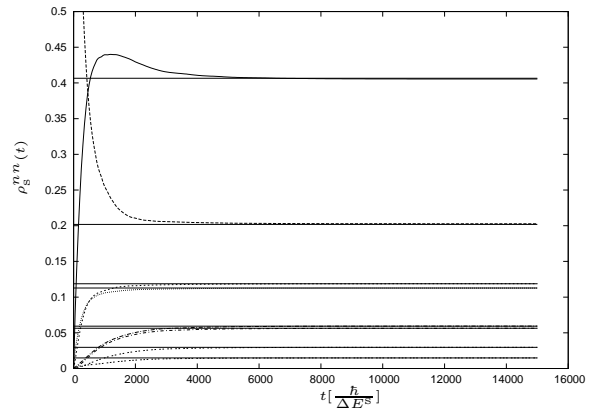}
\caption{\label{fig:System-Master-H} Time evolution of $\rho^{\text
    {s}}$ under a master equation with antiferromagnetic Heisenberg
  interaction. (compare Fig. \ref{fig:System-Sdyn-H}).} 
\end{figure}

We have solved the master equation of Sec.~\ref{sec:level24} with the
same values for the parameter $\lambda=0.4$ and $\kappa=0.001$ as for
the Schr\"odinger dynamics. The corresponding relaxation into
equilibrium is shown in Fig.~\ref{fig:System-Master-Z} for a random
coupling and in Fig.~\ref{fig:System-Master-H} for the
antiferromagnetic Heisenberg coupling. 

The horizontal lines denote, again, the equilibrium state, which the
spin-system should reach according to (\ref{eq:18}). The state
obtained via the master equation with only one spin coupled directly
to the bath approaches the same equilibrium state as the Schr\"odinger
dynamics. Also the off-diagonal elements are damped
away. Asymptotically both methods show, under the analyzed conditions,
the same behavior whereas the short-time behavior is difficult to
compare.

\begin{figure}[htbp]
\psfrag{T}{\tiny $k_{\text {B}} T[\Delta E]$}
\psfrag{lambda}{\tiny $\lambda[\Delta E^{\text {s}}]$}
\psfrag{exp}{\tiny $T^{\text {s}}$}
\psfrag{1}{\tiny $T^{\text {loc}}_1$}
\psfrag{2}{\tiny $T^{\text {loc}}_2$}
\psfrag{3}{\tiny $T^{\text {loc}}_3$}
\includegraphics[width=.3\textwidth,angle=270]{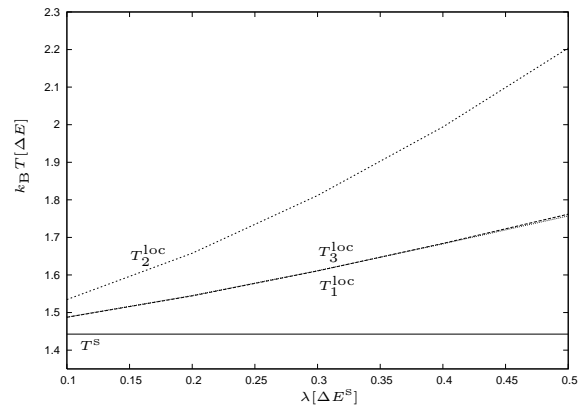}
\caption{\label{fig:Temp-Master-Z} Spectral-temperatures under
  a master equation with random interaction as function of
  $\lambda$. The solid line is the spectral-temperature $T^{\text
    {s}}$ of the total spin chain. The dashed line is the local temperature
  $T^{\text {loc}}_1$ of spin 1, the narrower dashed line $T^{\text
    {loc}}_2$ the one of spin 2 and the dotted line $T^{\text
    {loc}}_3$ the one of spin 3.}
\end{figure}

\begin{figure}[htbp]
\psfrag{T}{\tiny $k_{\text {B}} T[\Delta E]$}
\psfrag{lambda}{\tiny $\lambda[\Delta E^{\text {s}}]$}
\psfrag{s}{\tiny $T^{\text {s}}$}
\psfrag{1}{}
\psfrag{2}{\tiny $T^{\text {loc}}_2$}
\psfrag{3}{\tiny $T^{\text {loc}}_1=T^{\text {loc}}_3$}
\includegraphics[width=.3\textwidth,angle=270]{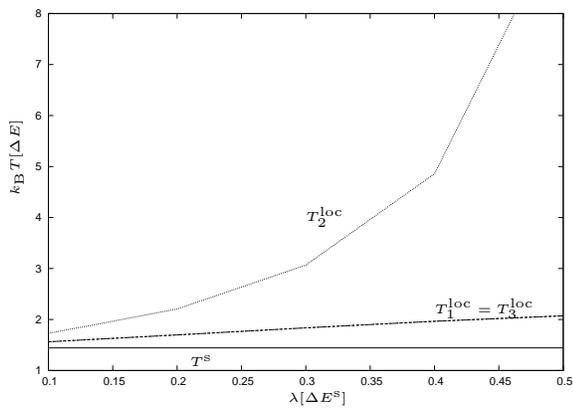}
\caption{\label{fig:Temp-Master-H} Spectral-temperatures under a
  master equation with antiferromagnetic Heisenberg interaction in
  dependence of $\lambda$ (compare Fig. \ref{fig:Temp-Master-Z}).}
\end{figure}

\emph{Global and local Temperatures:}
As for the Schr\"odinger dynamics we now study the global and local
states of the spin chain in dependence of the parameter $\lambda$. We
have calculated the spectral-temperature (\ref{eq:20}) for the whole
chain and for each individual spin (\ref{eq:24}). The results are
plotted in Fig.~\ref{fig:Temp-Master-Z} for a random interaction and
in Fig.~\ref{fig:Temp-Master-H} for a antiferromagnetic Heisenberg
interaction.


In both cases the temperature of the whole spin-chain $T^{\text {s}}$
reaches the same value as the temperature of the environment $T^{\text {e}} =
1.44$. For weak spin-spin coupling ($\lambda \ll 1$) the local
spectral-temperature of each spin is approximately the same as the one
of the whole chain. As for the Schr\"odinger dynamics with increasing
$\lambda$ the local spectral-temperature of each spin
rises. Especially the local temperature of the spin in the middle of
the chain, spin~2, increases more rapidly than those at the
boundaries. 

\section{\label{sec:analytic} Analytical Results}
\begin{figure}[htbp]
\psfrag{c}{\tiny $k_{\text {B}} T[\Delta E]$}
\psfrag{l}{\tiny $\lambda[\Delta E^{\text {s}}]$}
\psfrag{T1}{\tiny $T^{\text {loc}}_1$ and  $T^{\text {loc}}_3$}
\psfrag{T2}{\tiny $T^{\text {loc}}_2$}
\includegraphics[width=.45\textwidth]{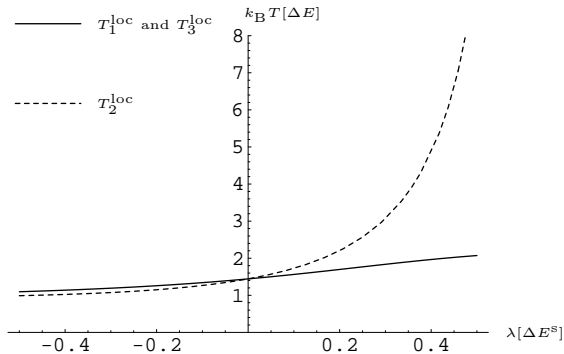}
\caption{\label{fig:Temp-analytic} Analytical results for local
  spectral-temperatures with the global inverse temperature $\beta=\ln
  2$ as function of $\lambda$ for ferromagnetic ($\lambda < 0$) and
  antiferromagnetic ($\lambda > 0$) Heisenberg coupling.}
\end{figure}

Finally we want to compare the numerical results shown before with the
analytic solutions for the Heisenberg coupled spin-chain. We
start from a density-matrix $\rho_{\text{S}}(\beta,\lambda)$ for the
whole spin-chain in a canonical state. Transforming
$\rho_{\text{S}}(\beta,\lambda)$ into the product basis, one can get
the reduced density matrices of the single spins
$\rho_\mu(\beta,\lambda)$. We have calculated the local
spectral-temperature for each spin using \eqref{eq:21} in dependence of
$\beta$ and $\lambda$. The result is shown in
Fig. \ref{fig:Temp-analytic} (with the same global inverse temperature
$\beta=\ln 2$ as for the numerical calculations) and is in good
agreement with our numerical results (compare with
Fig. \ref{fig:Temp-Schr-H} and Fig. \ref{fig:Temp-Master-H}, respectively).

\begin{figure}[htbp]
\psfrag{c}{\tiny $C$}%
\psfrag{l}{\tiny $\lambda[\Delta E^{\text {s}}]$}%
\includegraphics[width=.4\textwidth]{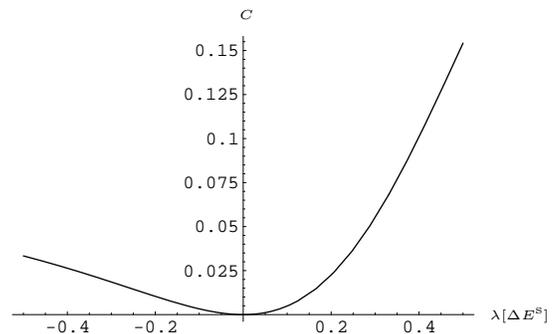}
\caption{\label{fig:C-ln2-analytic} Correlations C as function of
  $\lambda$ ($\beta=\ln 2$) for ferromagnetic ($\lambda < 0$) and
  antiferromagnetic ($\lambda > 0$) Heisenberg coupling.}
\end{figure}

To analyze the analytical solution for the correlation-measure
$C(\beta,\lambda)$  we have used again \eqref{eq:22} with the analytical
density matrices $\rho_{\text{S}}(\beta,\lambda)$ and
$\rho_\mu(\beta,\lambda)$. Fig. \ref{fig:C-ln2-analytic} shows
$C(\beta,\lambda)$ for the global inverse temperature $\beta=\ln 2$.
Again the analytical result is in very good accordance with our
numerical results (compare with Fig. \ref{fig:Corr-gl-H}). An
interesting point is, that the correlations $C$ also increase in the
ferromagnetic case although it is known that there exists no nearest
neighbor entanglement \cite{Arnesen2001}.

\begin{figure}[htbp]
\psfrag{c}{\tiny $C$}
\psfrag{b}{\tiny $\beta$}
\psfrag{l}{\tiny $\lambda[\Delta E^{\text {s}}]$}
\includegraphics[width=.5\textwidth]{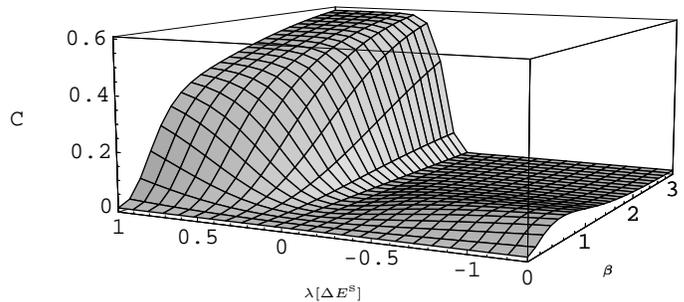}
\caption{\label{fig:C-analytic} Correlations $C$ as function of
  $\beta$ and $\lambda$ for ferromagnetic ($\lambda < 0$) and
  antiferromagnetic ($\lambda > 0$) Heisenberg coupling.}
\end{figure}

Fig. \ref{fig:C-analytic} shows $C(\beta,\lambda)$ as function of
$\beta$ and $\lambda$. $C$ increases with increasing $\beta$ and
$\lambda$ in the antiferromagnetic case as one would expect because of
the entanglement in the spin-chain for low temperatures. For the ferromagnetic
coupling one can see a bump depending on $\beta$ and $\lambda$. For
low temperatures (large $\beta$) $C$ decreases and vanish for
$T=0$.

Presently it is not quite clear why this bump occurs in the
ferromagnetic case for intermediate temperatures. It can be shown,
that the crest of the ferromagnetic bump is a function of $\beta$ and
$\lambda$. The behavior for high temperatures is the same for the
ferromagnetic as well as the antiferromagnetic case: $C$ is vanishing
because the local states and the global state will be a totally mixed
one. 

Note that $C$ is not an entanglement measure. $C$ checks whether
$\rho_{\text{S}}(\beta,\lambda)$ is factorisable by the reduced
density matrices $\rho_\mu(\beta,\lambda)$. Therefore $C \ge 0$ in the
ferromagnetic case for $T > 0$ indicates on one hand that the product basis
differs from the eigenbasis of the spin-chain \cite{Hartmann2004III}
(which is a pure quantum mechanical effect) and otherwise whether
a local thermodynamical description is possible.

\section{\label{sec:con} Summary and Conclusion}
The main motivation of this paper has been to analyze the thermodynamic
behavior of one part (``the system'') within a bipartite quantum
system subject to Schr\"odinger evolution. Such effective relaxation
to equilibrium is predicted to occur under rather general
conditions. Nevertheless deviations are not excluded. Here we have
been interested in (spatially) selective system-environment couplings
and in the effect of further partitioning of the system under
consideration.

For this purpose we have studied a spin-chain consisting of three
interacting spins coupled locally to a quantum environment.  Two different
spin-spin-coupling types have been examined: a random and a Heisenberg
coupling. For both we have solved the exact Schr\"odinger equation and
analyzed the reached equilibrium states. We showed that the
spin system relaxes into a thermal equilibrium state, which is in
accordance with the state one would expect from quantum
thermodynamics. Temporal fluctuations persist, though, and are a
result of the still comparatively small environment.

The spin chain always reaches the predicted state independently of
the internal coupling type and strength. Thus we can conclude, that
the internal structure of a small system as well as locality of the
coupling to a quantum environment does not effect the relaxation
behavior. The results also show that the overall state reached is a
canonical one and thus the spectral-temperature of the total
spin chain can be identified as the "real" (thermodynamic) temperature
of the system.

On the other hand, the spectral-temperature of each individual spin does
depend on the internal coupling strength, an effect which has been
traced back to increased correlations of each spin with its
neighbor(s). This causes an increase of the local entropy which
mimics a higher spectral-temperature. So our conclusion is that for
stronger internal couplings the local spin states are non-thermal
states \cite{Hartmann2005}, indicating that the global temperature
ceases to be available also locally.

For the antiferromagnetic Heisenberg spin chain, we have found that
single spins are approximately in thermal states up to coupling
strengths $\lambda \approx 1.5$. This is in agreement with the results in
\cite{Hartmann2005II}. Also the comparison with the analytical solution
shows no differences with the numerical results for the reached
equilibrium states.

A further motivation of this paper has been to compare the well-known
Markovian master equation approach with the exact Schr\"odinger
dynamics. We have found, that in all considered cases both approaches
lead to the same asymptotic result, i.\ e.\ equilibrium states of canonical
form. The analysis of the short time dynamics, e.\ g.\ the relaxation
times, etc., for the considered approaches should be an interesting
topic for future research. 

\begin{acknowledgments}
We thank H. Schmidt, M. Stollsteimer, F. Tonner and C. Kostoglou for
fruitful discussions. We thank the Deutsche Forschungsgemeinschaft for
financial support.
\end{acknowledgments}


\begin{thebibliography}{22}
\expandafter\ifx\csname natexlab\endcsname\relax\def\natexlab#1{#1}\fi
\expandafter\ifx\csname bibnamefont\endcsname\relax
  \def\bibnamefont#1{#1}\fi
\expandafter\ifx\csname bibfnamefont\endcsname\relax
  \def\bibfnamefont#1{#1}\fi
\expandafter\ifx\csname citenamefont\endcsname\relax
  \def\citenamefont#1{#1}\fi
\expandafter\ifx\csname url\endcsname\relax
  \def\url#1{\texttt{#1}}\fi
\expandafter\ifx\csname urlprefix\endcsname\relax\def\urlprefix{URL }\fi
\providecommand{\bibinfo}[2]{#2}
\providecommand{\eprint}[2][]{\url{#2}}

\bibitem[{\citenamefont{Neumann}(1929)}]{Neumann1929}
\bibinfo{author}{\bibfnamefont{J.~v.} \bibnamefont{Neumann}},
  \bibinfo{journal}{Z. Phys.} \textbf{\bibinfo{volume}{57}},
  \bibinfo{pages}{30} (\bibinfo{year}{1929}).

\bibitem[{\citenamefont{Lindblad}(1983)}]{Lindblad1983}
\bibinfo{author}{\bibfnamefont{G.}~\bibnamefont{Lindblad}},
  \emph{\bibinfo{title}{Non-equilibrium Entropy and Irreversibility}},
  vol.~\bibinfo{volume}{5} of \emph{\bibinfo{series}{Mathematical physics
  studies}} (\bibinfo{publisher}{Reidel}, \bibinfo{address}{Dordrecht},
  \bibinfo{year}{1983}).

\bibitem[{\citenamefont{Breuer and Petruccione}(2002)}]{Breuer}
\bibinfo{author}{\bibfnamefont{H.-P.} \bibnamefont{Breuer}} \bibnamefont{and}
  \bibinfo{author}{\bibfnamefont{F.}~\bibnamefont{Petruccione}},
  \emph{\bibinfo{title}{The {T}heory of {O}pen {Q}uantum {S}ystems}}
  (\bibinfo{publisher}{Oxford University Press}, \bibinfo{year}{2002}).

\bibitem[{\citenamefont{Weiss}(1999)}]{Weiss1999}
\bibinfo{author}{\bibfnamefont{U.}~\bibnamefont{Weiss}},
  \emph{\bibinfo{title}{Quantum Dissipative Systems}}
  (\bibinfo{publisher}{World Scientific}, \bibinfo{address}{Singapore, New
  Jersey, London, Hong Kong}, \bibinfo{year}{1999}), \bibinfo{edition}{2nd} ed.

\bibitem[{\citenamefont{Kossakowski}(1972)}]{Kossakowski1972}
\bibinfo{author}{\bibfnamefont{A.}~\bibnamefont{Kossakowski}},
  \bibinfo{journal}{Bull. Acad. Polon. Sci. S\'er. Sci. Math. Astronom. Phys.}
  \textbf{\bibinfo{volume}{20}}, \bibinfo{pages}{1021} (\bibinfo{year}{1972}).

\bibitem[{\citenamefont{Lindblad}(1975)}]{Lindblad1975}
\bibinfo{author}{\bibfnamefont{G.}~\bibnamefont{Lindblad}},
  \bibinfo{journal}{Commun. Math. Phys.} \textbf{\bibinfo{volume}{40}},
  \bibinfo{pages}{147} (\bibinfo{year}{1975}).

\bibitem[{\citenamefont{Gorini et~al.}(1976)\citenamefont{Gorini, Kossakowski,
  and Sudarshan}}]{Gorini1976}
\bibinfo{author}{\bibfnamefont{V.}~\bibnamefont{Gorini}},
  \bibinfo{author}{\bibfnamefont{A.}~\bibnamefont{Kossakowski}},
  \bibnamefont{and} \bibinfo{author}{\bibfnamefont{E.~C.~G.}
  \bibnamefont{Sudarshan}}, \bibinfo{journal}{J. Mathematical Phys.}
  \textbf{\bibinfo{volume}{17}}, \bibinfo{pages}{821} (\bibinfo{year}{1976}).

\bibitem[{\citenamefont{Gemmer et~al.}(2001)\citenamefont{Gemmer, Otte, and
  Mahler}}]{GemmerOtte2001}
\bibinfo{author}{\bibfnamefont{J.}~\bibnamefont{Gemmer}},
  \bibinfo{author}{\bibfnamefont{A.}~\bibnamefont{Otte}}, \bibnamefont{and}
  \bibinfo{author}{\bibfnamefont{G.}~\bibnamefont{Mahler}},
  \bibinfo{journal}{Phys.\ Rev.\ Lett.} \textbf{\bibinfo{volume}{86}},
  \bibinfo{pages}{1927} (\bibinfo{year}{2001}).

\bibitem[{\citenamefont{Gemmer et~al.}(2005)\citenamefont{Gemmer, Michel, and
  Mahler}}]{GeMiMa2004}
\bibinfo{author}{\bibfnamefont{J.}~\bibnamefont{Gemmer}},
  \bibinfo{author}{\bibfnamefont{M.}~\bibnamefont{Michel}}, \bibnamefont{and}
  \bibinfo{author}{\bibfnamefont{G.}~\bibnamefont{Mahler}},
  \emph{\bibinfo{title}{Quantum {T}hermodynamcis - {E}mergence of
  {T}hermodynamic {B}ehavior within {C}omposite {Q}uantum {S}ystems}},
  \bibinfo{number}{LNP 657} (\bibinfo{publisher}{Springer LNP, Berlin, New
  York}, \bibinfo{year}{2005}).

\bibitem[{\citenamefont{Borowski et~al.}(2003)\citenamefont{Borowski, Gemmer,
  and Mahler}}]{BorowskiGemmer2003}
\bibinfo{author}{\bibfnamefont{P.}~\bibnamefont{Borowski}},
  \bibinfo{author}{\bibfnamefont{J.}~\bibnamefont{Gemmer}}, \bibnamefont{and}
  \bibinfo{author}{\bibfnamefont{G.}~\bibnamefont{Mahler}},
  \bibinfo{journal}{Euro. Phys. J. B} \textbf{\bibinfo{volume}{35}},
  \bibinfo{pages}{255} (\bibinfo{year}{2003}).

\bibitem[{\citenamefont{Hartmann
  et~al.}(2004{\natexlab{a}})\citenamefont{Hartmann, Mahler, and
  Hess}}]{Hartmann2004III}
\bibinfo{author}{\bibfnamefont{M.}~\bibnamefont{Hartmann}},
  \bibinfo{author}{\bibfnamefont{G.}~\bibnamefont{Mahler}}, \bibnamefont{and}
  \bibinfo{author}{\bibfnamefont{.}~\bibnamefont{Hess}},
  \bibinfo{journal}{Phys. Rev. Lett.} \textbf{\bibinfo{volume}{93}},
  \bibinfo{pages}{080402} (\bibinfo{year}{2004}{\natexlab{a}}).

\bibitem[{\citenamefont{Hartmann
  et~al.}(2004{\natexlab{b}})\citenamefont{Hartmann, Mahler, and
  Hess}}]{Hartmann2004IV}
\bibinfo{author}{\bibfnamefont{M.}~\bibnamefont{Hartmann}},
  \bibinfo{author}{\bibfnamefont{G.}~\bibnamefont{Mahler}}, \bibnamefont{and}
  \bibinfo{author}{\bibfnamefont{O.}~\bibnamefont{Hess}},
  \bibinfo{journal}{Phys. Rev. E} \textbf{\bibinfo{volume}{70}},
  \bibinfo{pages}{066148} (\bibinfo{year}{2004}{\natexlab{b}}).

\bibitem[{\citenamefont{Michel et~al.}(2003)\citenamefont{Michel, Hartmann,
  Gemmer, and Mahler}}]{Michel2003}
\bibinfo{author}{\bibfnamefont{M.}~\bibnamefont{Michel}},
  \bibinfo{author}{\bibfnamefont{M.}~\bibnamefont{Hartmann}},
  \bibinfo{author}{\bibfnamefont{J.}~\bibnamefont{Gemmer}}, \bibnamefont{and}
  \bibinfo{author}{\bibfnamefont{G.}~\bibnamefont{Mahler}},
  \bibinfo{journal}{Euro. Phys. J. B} \textbf{\bibinfo{volume}{34}},
  \bibinfo{pages}{325} (\bibinfo{year}{2003}).

\bibitem[{\citenamefont{Gemmer and Mahler}(2003)}]{Gemmer2003}
\bibinfo{author}{\bibfnamefont{J.}~\bibnamefont{Gemmer}} \bibnamefont{and}
  \bibinfo{author}{\bibfnamefont{G.}~\bibnamefont{Mahler}},
  \bibinfo{journal}{Euro. Phys. J. B} \textbf{\bibinfo{volume}{31}},
  \bibinfo{pages}{249} (\bibinfo{year}{2003}).

\bibitem[{\citenamefont{Mehta}(1991)}]{Mehta1991}
\bibinfo{author}{\bibfnamefont{M.~L.} \bibnamefont{Mehta}},
  \emph{\bibinfo{title}{Random {M}atrices}} (\bibinfo{publisher}{Academic
  Press, Boston USA}, \bibinfo{year}{1991}).

\bibitem[{\citenamefont{Lindblad}(1976)}]{Lindblad1976}
\bibinfo{author}{\bibfnamefont{G.}~\bibnamefont{Lindblad}},
  \bibinfo{journal}{Comm. Math. Phys.} \textbf{\bibinfo{volume}{48}},
  \bibinfo{pages}{119} (\bibinfo{year}{1976}).

\bibitem[{\citenamefont{Saito et~al.}(2000)\citenamefont{Saito, Takesue, and
  Miyashita}}]{Saito2000}
\bibinfo{author}{\bibfnamefont{K.}~\bibnamefont{Saito}},
  \bibinfo{author}{\bibfnamefont{S.}~\bibnamefont{Takesue}}, \bibnamefont{and}
  \bibinfo{author}{\bibfnamefont{S.}~\bibnamefont{Miyashita}},
  \bibinfo{journal}{Phys. Rev. E} \textbf{\bibinfo{volume}{61}},
  \bibinfo{pages}{2397} (\bibinfo{year}{2000}).

\bibitem[{\citenamefont{$\breve{\text C}$\'apek and
  Brav\'ik}(2001)}]{Capek2000}
\bibinfo{author}{\bibfnamefont{V.}~\bibnamefont{$\breve{\text C}$\'apek}}
  \bibnamefont{and} \bibinfo{author}{\bibfnamefont{I.}~\bibnamefont{Brav\'ik}},
  \bibinfo{journal}{Physica A} \textbf{\bibinfo{volume}{294}},
  \bibinfo{pages}{388 } (\bibinfo{year}{2001}).

\bibitem[{\citenamefont{Landau and Lifschitz}(1971)}]{LL5}
\bibinfo{author}{\bibfnamefont{L.~D.} \bibnamefont{Landau}} \bibnamefont{and}
  \bibinfo{author}{\bibfnamefont{E.~M.} \bibnamefont{Lifschitz}},
  \emph{\bibinfo{title}{Statistische Physik}} (\bibinfo{publisher}{Akademie
  Verlag Berlin}, \bibinfo{address}{Berlin}, \bibinfo{year}{1971}),
  \bibinfo{edition}{3rd} ed.

\bibitem[{\citenamefont{Arnesen et~al.}(2001)\citenamefont{Arnesen, Bose, and
  Vedral}}]{Arnesen2001}
\bibinfo{author}{\bibfnamefont{M.~C.} \bibnamefont{Arnesen}},
  \bibinfo{author}{\bibfnamefont{S.}~\bibnamefont{Bose}}, \bibnamefont{and}
  \bibinfo{author}{\bibfnamefont{V.}~\bibnamefont{Vedral}},
  \bibinfo{journal}{Phys. Rev. Lett.} \textbf{\bibinfo{volume}{87}},
  \bibinfo{pages}{017901} (\bibinfo{year}{2001}).

\bibitem[{\citenamefont{Hartmann
  et~al.}(2005{\natexlab{a}})\citenamefont{Hartmann, Mahler, and
  Hess}}]{Hartmann2005}
\bibinfo{author}{\bibfnamefont{M.}~\bibnamefont{Hartmann}},
  \bibinfo{author}{\bibfnamefont{G.}~\bibnamefont{Mahler}}, \bibnamefont{and}
  \bibinfo{author}{\bibfnamefont{O.}~\bibnamefont{Hess}},
  \bibinfo{journal}{cond-mat/0410526}  (\bibinfo{year}{2005}{\natexlab{a}}).

\bibitem[{\citenamefont{Hartmann
  et~al.}(2005{\natexlab{b}})\citenamefont{Hartmann, Mahler, and
  Hess}}]{Hartmann2005II}
\bibinfo{author}{\bibfnamefont{M.}~\bibnamefont{Hartmann}},
  \bibinfo{author}{\bibfnamefont{G.}~\bibnamefont{Mahler}}, \bibnamefont{and}
  \bibinfo{author}{\bibfnamefont{O.}~\bibnamefont{Hess}}, \bibinfo{journal}{J.
  Phys. Soc. Jpn.} \textbf{\bibinfo{volume}{74}}
  (\bibinfo{year}{2005}{\natexlab{b}}).

\end{thebibliography}
\end{document}